\documentclass[letter]{aa} % for the letters 
\usepackage{graphicx}
\usepackage{txfonts}
\newcommand{\mygi}{MyGIsFOS}
\newcommand{\vr}{\ensuremath{V_\mathrm{r}}}

\newcommand{\logg}{\ensuremath{\log\,g}}
\def\teff{$T\rm_{eff}$}
\newcommand{\kms}{$\rm km s ^{-1}$}

\begin{document} 

\title{High speed stars: III.\\ Detailed abundances and binary nature of the extreme speed star GHS143
\thanks{Based on observations made with UVES at VLT 111.24J1.001 and 112.25EH.001.}
\thanks{The lines investigated are on an on-line table at CDS.}
}
\titlerunning{High speed stars III}

\author{
E.~Caffau    \inst{1} \and
P.~Bonifacio \inst{1} \and
L.~Monaco \inst{2} \and
L.~Sbordone \inst{3} \and      
M.~Spite \inst{1} \and
P.~Fran\c{c}ois \inst{4,5} \and
P.~Panuzzo \inst{1} \and      
P.~Sartoretti \inst{1} \and    
L.~Chemin \inst{6} \and        
F.~Th\'evenin \inst{7} \and
A.~Mucciarelli \inst{8}
}

\institute{GEPI, Observatoire de Paris, Universit\'{e} PSL, CNRS,  5 Place Jules Janssen, 92190 Meudon, France
%              \email{}
\and
Universidad Andres Bello, Facultad de Ciencias Exactas, Departamento de Ciencias F{\'\i}sicas - Instituto de Astrof{\'\i}sica, Autopista
Concepci{\'o}n-Talcahuano, 7100, Talcahuano, Chile
\and
European Southern Observatory, Casilla 19001, Santiago, Chile
\and
GEPI, Observatoire de Paris, Universit\'{e} PSL, CNRS, 77 Av. Dendert-Rocheau, 75014 Paris, France
\and
UPJV, Universit\'e de Picardie Jules Verne, 33 rue St Leu, 80080 Amiens, France
\and
Universidad Andres Bello, Facultad de Ciencias Exactas, Departamento de Ciencias F\'isicas - Instituto de Astrof\'isica, Fernandez Concha 700, Las Condes, Santiago, Chile
\and
Universit\'e de Nice Sophia-Antipolis, CNRS, Observatoire de la C\^ote d'Azur, Laboratoire Lagrange, BP 4229, F-06304 Nice, France
\and
Dipartimento di Fisica e Astronomia, Universit\`a degli Studi di Bologna, Via Gobetti 93/2, I-40129 Bologna, Italy
}

   \date{Received September 15, 1996; accepted March 16, 1997}

  \abstract
{The Gaia satellite has provided the community with three releases containing astrometrical and photometric data as well as by products, 
such as stellar parameters and variability indicators.
}
{By selecting in the Gaia database, one can select stars with the  requested characteristics, such as high speed.
At present any selection is based on available Gaia releases including a subset of the observations.
This, for some stars, can show some limitations,
for example there is  still not a sufficient number of observations to detect  binarity. 
}
{We investigated a star selected in Gaia EDR3 for its high speed that  appears unbound to the Galaxy.
We requested high-quality spectra to derive more information on the star.
}
{From the spectroscopic investigation we confirm the low metallicity content of the star, and
we derive a detailed chemical composition.
The star is poor in carbon and very rich in oxygen:   [(C+N+O)/Fe]=+0.65.
From the two spectra observed we conclude that the star is in a binary system
and from the investigation of the ionisation balance we derive that the star is closer
than implied by the  Gaia DR3 parallax, and thus has a  a lower intrinsic luminosity.}
{The star is probably still unbound, but there is the possibility that it is bound to the Galaxy.
Its low carbon abundance suggests that the star was formed in  a dwarf galaxy.}

   \maketitle
%
%-------------------------------------------------------------------
\section{Introduction\label{intro}}

In an ongoing investigation, we are trying to characterise the population
of stars with high speed with respect to the Sun.
In \citet[][hereafter Paper\,I]{ghs1}, we selected in Gaia\,Data Release 2 (DR2) \citep{gaiadr2} a sample of 72 stars for their 
high transverse velocity
($\rm V_{trans}>500$\,\kms). In the sample, we highlighted a few apparently young stars,
according to their position in the Gaia colour--absolute magnitude diagram.
The sample was further 
discussed in \citet[][hereafter Paper\,II]{ghs2}, where two other  samples, selected with the same  criterion on $\rm V_{trans}$
in Gaia\,EDR3 \citep{gaiadr3}, are analysed.
In Paper\,II other seemingly young metal-poor stars are detected and discussed.
In particular, GHS143 (Gaia\,DR3\,6632370485122299776) is a metal-poor evolved star of G magnitude 13.06 
and of apparent young age, characterised by   extreme kinematics.
Assuming the parallax, proper motions, and radial
velocity in Gaia\,DR3 \citep{gaiadr3}, it can be seen that    GHS143  is not bound to the Galaxy, but is falling into it.
We requested high-resolution spectra of this star to investigate if the stellar parameters we derived from a high-resolution spectrum
are consistent with the Gaia\,DR3 parallax and photometry, and also to check for possible
radial velocity variations. 
In this paper we use these spectra to derive a complete
chemical inventory of this star, derive the uncertainties,
and discuss the possibility that the star is bound to the Galaxy.
  
%%%%%%%%%%%%%%%%%%%%%%%%%%%%%%%%%%%%%%%%%%%%%%%%%%%%%%%%%%%%%%%%%%%%%%%% 
\section{Observations} 

\begin{figure}
\centering
\includegraphics[width=\hsize,clip=true]{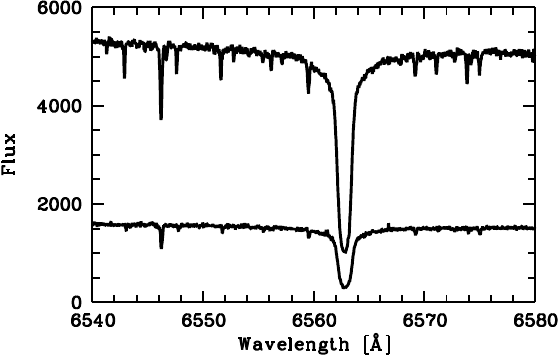}
\caption{Observed spectra in the H$\alpha$ region.
The spectrum with higher flux is from November 2023; the lower flux is from August 2023.}
\label{fig:obs}
\end{figure}

Two UVES \citep{uves} spectra have been secured for this star.
In the ESO programme 111.22EH.001 the star was observed on August 20, 2023,
in the setting DIC2\,437+760 (wavelength ranges 373--499 and 565--946\,nm),
with slit 0\farcs{4}\ (resolving power 90\,000) in the  blue arm and 0\farcs{3}\ (resolving power 110\,000) in the red arm.
In the ESO programme 0112.25EH.001 the star was observed on November 11, 2023,
with the same setting and with slit 0\farcs{5}\ (resolving power 75\,000) in the  blue arm and in the red arm.
Both observations  were graded `A'.
The signal-to-noise ratio of the November spectrum is better than the August spectrum 
(S/N of 53 and 35 at 400\,nm; 79 and 49 at 498\,nm; 97 and 72 at 640\,nm; 86 and 67 at 838\,nm).
We reduced the spectra using the ESO UVES pipeline.
In Fig.\,\ref{fig:obs} the two UVES spectra in the $\rm H\alpha$ region are shown.

%%%%%%%%%%%%%%%%%%%%%%%%%%%%%%%%%%%%%%%%%%%%%%%%%%%%%%%%%%%%%%
%%%%%%%%%%%%%%%%%%%%%%%%%%%%%%%%%%%%%%%%%%%%%%%%%%%%%%%%%%%%%%

%%%%%%%%%%%%%%%%%%%%%%%%%%%%%%%%%%%%%%%%%%
\section{Analysis}

%%%%%%%%%%%%%%%%%%%%%%%%%%
\subsection{Radial velocity} \label{sec:vr}

We used template matching to measure the radial velocities from the
observed spectra. As a template we used a synthetic spectrum with
the parameters derived in \citet{ghs2}.
The spectra were corrected for the barycentric Earth velocity using
the value in the descriptor {\tt ESO.QC.VRAD.BARYCOR} available
in the reduced spectrum.
We measured a radial velocity for each
of the spectra collected in the three UVES detectors (one in the blue arm
and two in the red arm). We used the following wavelength
ranges: 376\,nm--497\,nm for the blue arm, 570\,nm--676\,nm for the lower detector
of the red arm, and 800\,nm--900\,nm for the upper detector of the red arm.
We then adopted the mean of these three measurements as the  radial velocity
and  the standard deviation of the three values as an estimate of the
error on the radial velocity.
From the UVES spectrum of August 2023 we derived a \vr\ of $22.71\pm 0.30$\,\kms\ and
from the spectrum of   November 2023 a \vr\ of $25.58\pm 0.40$\,\kms,
to be compared to the Gaia\,DR3 \vr\ of $17.5\pm 1.64$\,\kms.
From these three values it is clear that the star displays radial velocity
variations that make it likely that the star belongs to a multiple system.
We add that the radial velocity measured on the low-resolution spectrum
discussed in Paper\,II is $-27.7\pm 4.0$\,\kms, where the error estimate
is purely statistical. The spectrum was taken with FORS, which  is subject to flexures, 
and the systematic error in radial velocity is 48.7\,\kms, estimated as in \citet{PristineXI}. 
Even so, this measurement, at face value, supports the existence of radial
velocity variations in this star.
A monitoring of its radial velocity, beyond what is available from 
the epoch radial velocities that are provided by Gaia, is
strongly encouraged.

%%%%%%%%%%%%%%%%%%%%%%%%%%
\subsection{Stellar parameters} \label{sec:param}

The star was analysed in Paper\,II and the stellar parameters adopted were
\teff=5159\,K, \logg=1.8\,dex, a microturbulence $\xi=1.96$\,\kms;\ and  an iron abundance of $\rm [Fe/H]=-1.74$\,dex was then derived. 
When adopting these stellar parameters to analyse the UVES spectra, we obtain   Fe abundances in very
good agreement ($\rm [Fe/H]=-1.86\pm 0.10$\,dex and $\rm [Fe/H]=-1.85\pm 0.10$\,dex from the two spectra),
and in both cases a good Fe ionisation balance.

We allowed \mygi\ to derive the parameters and we obtained for \teff, \logg, $\xi$, and [Fe/H]:
$4988\pm 96$, $1.37\pm 0.05$, $1.59\pm 0.07$, and $-1.92\pm 0.10$ from the spectrum of August and
$4956\pm 93$, $1.40\pm 0.07$, $1.54\pm 0.08$, and $-1.95\pm 0.12$ from the spectrum of November.
The two spectra provide very coherent results.
We used the calibration suggested by \citet{frebel13} to bring the effective temperature derived by the excitation 
on the photometric scale. 
With this calibration we derived 5159 and 5127\,K, respectively, in excellent agreement with the value derived from the Gaia\,DR3 photometry
and applied in Paper\,II.
By using the calibration by \citet{mucciarelli20} to derive \teff\ on the photometric scale of \citet{GHB09},
we derived 5136 and 5112\,K, respectively, with an uncertainty of 130\,K.
We adopted \teff = 5160\,K.

\begin{figure}
\centering
\includegraphics[width=\hsize,clip=true]{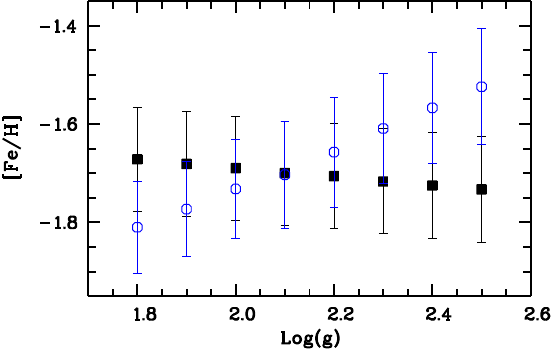}
\caption{[Fe/H] derived from \ion{Fe}{i} (filled black squares) and corrected by NLTE and from \ion{Fe}{ii} (blue open circles) lines
vs \logg.}
\label{fig:loggfe}
\end{figure}

To derive the stellar parameters we focused on the spectrum from November 2023, which has a higher flux, and to derive the uncertainties we used the August 2023 spectrum.
We are aware that the Fe abundance derived from \ion{Fe}{i} lines is affected by 
departure from local termodymanycal equilibrium (hereafter NLTE, the local termodymanycal equilibrium shall be referred to as LTE)
and forcing the ionisation equilibrium does not take into account the NLTE effects.
The NLTE correction we expect for this star is about 0.1\,dex. 
We selected the \ion{Fe}{i} lines retained by \mygi\ and also available from the  web site of MPIA,\footnote{\url{https://nlte.mpia.de/gui-siuAC_secE.php}} 
and with these 39 \ion{Fe}{i} lines (providing $\rm [Fe/H]=-1.76$) we derived the NLTE corrections \citep{bergemann12fe}.
For a sample of 30 \ion{Fe}{i} lines with NLTE corrections from \citep{bergemann12fe}, we verified the NLTE corrections as
$\rm \langle 3D \rangle_{\rm NLTE}-1D_{\rm LTE}$ in \citet{amarsi16} and, 
with exactly the same stellar parameters, the average difference is 0.02\,dex.
We gave as input to \mygi\ several \logg\ values (see Fig.\,\ref{fig:loggfe}) with fixed \teff;\  
we derived the best agreement in [Fe/H] from the  \ion{Fe}{i} and \ion{Fe}{ii} lines for $\logg = 2.1$\,dex. 
The NLTE correction is of 0.11\,dex for $\logg =2.1$ and spans values from 0.08\,dex for the highest surface gravity to 0.13\,dex for the lowest.
These corrections are applied in Fig.\,\ref{fig:loggfe}.
With fixed effective temperature and assuming LTE, the iron
ionisation balance implies \hbox{\logg\,=\,1.8}, as expected, lower
than the value implied by the NLTE iron abundances.

We fixed \teff\ and \logg\ to derive the microturbulence and derived $1.65\pm 0.07$\,\kms,
changing very little by changing \logg\ (a change of 0.02\,\kms\ for a variation of 0.4\,dex in \logg).
The microturbulence derived by using the calibration of \citet{mashonkina17},
with \teff=5159\,K and \logg=2.1\,dex, provided 1.83\,\kms.

By assuming \teff=5159\,K, \logg=2.1\,dex, and $\xi=1.65$\,\kms, as derived by \mygi,
we derived [Fe/H] of $-1.81\pm 0.11$ and $-1.70\pm 0.11$, when derived from \ion{Fe}{i} and \ion{Fe}{ii} lines, 
respectively, in perfect agreement if we expect an NLTE correction of 0.11\,dex\ \citep{bergemann12fe} on [Fe/H] from Fe neutral lines.
In Fig. \ref{fig:loggfe} we show how the iron abundance derived from \ion{Fe}{i} and \ion{Fe}{ii}
lines changes, as a function of the adopted log g. 
All values of surface gravity in the range 1.8--2.5 are acceptable, 
considering the involved uncertainties. 

%%%%%%%%%%%%%%%%%%%%%%%%%%%%%%%%%%%%%%%%%%%%%%%%%%%%%%%%%%%%%%
%%%%%%%%%%%%%%%%%%%%%%%%%%%%%%%%%%%%%%%%%%%%%%%%%%%%%%%%%%%%%%

%%%%%%%%%%%%%%%%%%%%%%%%%%%%%%%
\subsection{Stellar mass} \label{sec:mass}

\begin{figure}
\centering
\includegraphics[width=\hsize,clip=true]{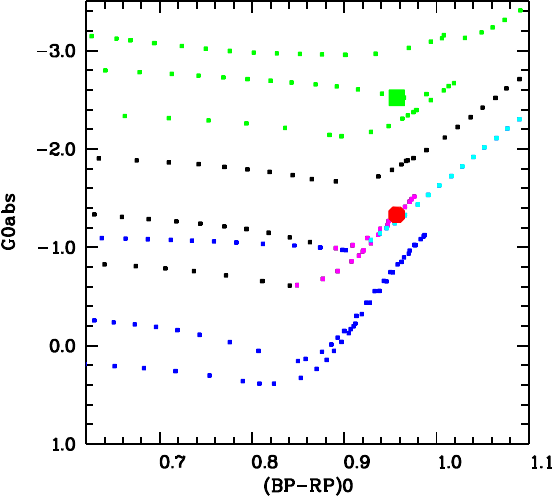}
\caption{GHS143 (red circle) in the colour magnitude diagram compared to two isochrones at metallicity $-1.65$
highlighting the ranges for possible solutions:
 500\,Ma (black   and pink, the evolution stages for the first two solutions)
and  1\,Ga (dark blue   and light blue, the evolution stage for the third solution).
The solution with the Gaia\,DR3 parallax (large green square) adopted in Paper\,II is compared to the isochrone with an age of 200\,Ma
(green points).}
\label{fig:iso}
\end{figure}

In Paper\,II, we adopted the parallax provided in Gaia\,DR3 corrected by the zero-point \citep{lindegren21}, and
we derived a mass of $\rm 3.1\, M_\odot$ or $\rm 3.8\, M_\odot$.
According to the Fe ionisation balance, from the UVES spectra, we favour a higher \logg\ that would
imply that the star has a lower intrinsic luminosity, so it has to be closer with a larger parallax.
In spite of the uncertainties that plague the spectroscopic surface gravity determination,
we believe that in this case it is more reliable than that derived from
the parallax. Since our radial velocity measurements
imply that the star is a binary, we expect that its astrometric measurements
also contain   a component due to the orbital motion. 
The astrometric data should then be processed using one of the astrometric
binary processing pipelines of Gaia \citep{Halbwachs2023,Holl2023} that
would result in a parallax  different
from that available in Gaia\,DR3, which was obtained treating the star as
a single star. As  mentioned above, the spectroscopic surface gravity implies
that this parallax should be larger than that in Gaia\,DR3.
With the adopted parameters (\teff=5159, \logg=2.1, [Fe/H]=--1.8), we derived these possible masses and ages (see Fig.\,\ref{fig:iso}):
\begin{itemize}
\item $\rm M=2.3 M_\odot$ and age of 493\,Ma (3, RGB, red giant branch, or the quick stage of red giant for intermediate+massive stars\footnote{\url{http://stev.oapd.inaf.it/cmd_3.1/faq.html}}); 
\item $\rm M=2.3 M_\odot$ and age of 504\,Ma (4, CHEB, core He-burning for low mass stars, or the very initial stage of CHeB for intermediate+massive stars);
\item $\rm M=1.9 M_\odot$ and age of 904\,Ma (7, EAGB, the early asymptotic giant branch).
\end{itemize}
In the most extreme case, with an adopted \logg\ of 2.5\,dex, we derived the following:
\begin{itemize}
\item $\rm M=1.4 M_\odot$ and age of 2\,Ga (3); 
\item $\rm M=0.8 M_\odot$ and age of 8.4\,Ga (7).
\end{itemize}

%%%%%%%%%%%%%%%%%%%%%%%%%%%%%%%%%%%%%%%%%%%%%%%%%%%%%%%%%%%%%%
%%%%%%%%%%%%%%%%%%%%%%%%%%%%%%%%%%%%%%%%%%%%%%%%%%%%%%%%%%%%%%

%%%%%%%%%%%%%%%%%%%%%%%%%%%%%%%
\subsection{Kinematics} \label{sec:kine}

The fact that the star is a spectroscopic binary implies that
the Gaia\,DR3 parallax may be incorrect. 
The Gaia\,DR3 parallax was derived assuming
that GHS143 is a single star.
The quality of the UVES
spectra allowed us to put strong constraints on the surface gravity
of the star, based on the iron ionisation equilibrium, for which we took into
account the NLTE effects. Our preferred gravity is $\logg = 2.1$;
taking into account the errors on the \ion{Fe}{i} and \ion{Fe}{ii} abundances,
any surface gravity in the interval 1.8 -- 2.5 is consistent with the
observations. If we turn around the Stefan--Boltzmann equation and associate
a parallax with each  surface gravity,  this translates into parallaxes from 0.098\,mas
to 0.222\,mas. We investigated how the dynamics of
the star changes for various parallaxes in this range.
For each parallax we proceeded as in Paper\,II; we employed
the {\tt galpy} code and the MWPotential2014 Galactic potential \citep{bovy15}
and the same assumptions on the solar position and motion.
For each parallax we considered the astrometric covariance matrix
and used the  {\tt Pyia} code \citep[][]{pyia18} to produce a random realisation
of the stellar kinematic data. For each parallax we extracted 1000 realisations
and used them as input to {\tt galpy} to evaluate the dynamical status of the star.

For all parallaxes smaller than 0.168\,mas, corresponding to log g = 2.24,
the star is unbound, as derived by \citet{ghs2} from the Gaia\,DR3 parallax.
For a parallax of 0.168\,mas the star is partially unbound, in the sense
that it is unbound for 609 realisations out of 1000. For larger parallaxes
the star becomes bound, albeit with a large apocentre, in excess of 30\,kpc
with our adopted Galactic potential.
We conclude that the boundary between being bound and unbound is 
around log g = 2.2, higher surface gravities make the star bound, while
lower values make it unbound, in the adopted Galactic potential.

%%%%%%%%%%%%%%%%%%%%%%%%%%%%%%%
\subsection{Abundances}\label{secabbo}

\begin{figure}
\centering
\includegraphics[width=\hsize,clip=true]{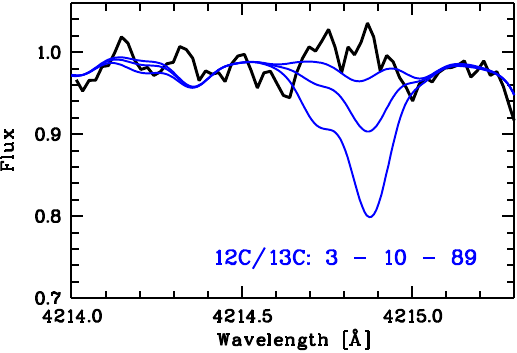}
\caption{Observed spectra (solid black) compared to theoretical synthesis in the G-band range where $\rm ^{13}C$ line are expected.}
\label{fig:ciso}
\end{figure}

The chemical investigation provided in Table\,\ref{tab:abbo} is from the spectrum observed in November 2023. 
The star is poor in carbon, but rich in nitrogen and oxygen.
The C abundance was derived by line profile fitting of the G-band at about 428\,nm.
We did not detect any $\rm ^{13}C$ (see Fig.\,\ref{fig:ciso}), and we concluded that the $\rm ^{12}C/^{13}C$ is not higher than solar.
We investigated the CN band at 383\,nm and, fixing the C abundance, we derived N abundance of A(N)=6.83.
Oxygen was derived from the [OI] line at 630\,nm and from the triplet at 777\,nm.
The lines of the triplet are affected by  non-negligible NLTE effects, while the forbidden line forms in conditions close to LTE.
By correcting the O abundance for NLTE effects \citep{sitnova13_o} the four \ion{O}{i} lines are in very close agreement, 
and in this case we derived [(C+N+O)/Fe]=0.65.

%%%%%%%%%%%%%%%%%%%%%%%%%%%%%%%%%%%%%%%%%%%%%%%%%%%%%%%%%%%%%%
%%%%%%%%%%%%%%%%%%%%%%%%%%%%%%%%%%%%%%%%%%%%%%%%%%%%%%%%%%%%%%

The star is enhanced in $\alpha$ elements.
From ten \ion{Mg}{i} lines we derive $\rm [Mg/Fe]=+0.52$ (see Fig.\,\ref{fig:alphafe}).
The NLTE corrections for Fe and Mg are comparable \citep[+0.08\,dex for the four \ion{Mg}{i} lines  used here;][]{bergemann17},
so the [Mg/Fe] ratio, taking into account the NLTE corrections, is close to the LTE value.
From 15 lines of \ion{Si}{i} we derived $\rm [Si/Fe]=+0.46$ and from one \ion{Si}{ii} line $\rm [Si/Fe]=+0.35$.
For the four lines the NLTE correction is small \citep[$-0.03$\,dex,][]{bergemann13}.
From 23 \ion{Ca}{i} lines we derived $\rm [Ca/Fe]=+0.46$ (see Fig.\,\ref{fig:alphafe}).
Of the \ion{Ca}{i} lines investigated, 11 provided a NLTE correction of +0.09\,dex \citep{mashonkina07}, similar to the NLTE correction of iron.

The A(Ti) is derived from \ion{Ti}{i} and \ion{Ti}{ii} features (23 and 32 lines, respectively), 
providing   an enhancement in [Ti/Fe]
($\rm [Ti/Fe]=0.40\pm 0.17$ and $\rm [Ti/Fe]=0.41\pm 0.13$, respectively). The 
Ti abundance derived from the  \ion{Ti}{ii} lines is close to the LTE condition \citep[see][]{sitnova20}, so this value should be preferred.

Sodium is derived from four \ion{Na}{i} lines (498.2, 616.0, 818.3, and 819.5\,nm) and provided $\rm [Na/Fe]=0.13\pm 0.23$ (see Fig.\,\ref{fig:alphafe}).
According to \citet{takeda03} the three reddest lines are affected by   NLTE corrections that reduce the Na abundance by
about $-0.2$\,dex and decrease the line-to-line scatter.
We analysed the strong \ion{Al}{i} line at 396\,nm. 
The NLTE correction for this line is 0.6\,dex, according to \citet{andrievsky08}.
The \ion{K}{i} line at 769.8\,nm provided $\rm [K/H]= -0.87$\,dex, so a strong K enhancement
($\rm [K/Fe] = 0.93$\,dex). Taking into account the NLTE correction provided by \citet{reggiani19},
we obtained a NLTE [K/Fe] ratio of $+0.41$\,dex, making the star still rich in K, but not so extreme.

We derived abundance values from neutral and single ionised V, Cr, and Mn, and the abundances derived are consistent
with that derived from \ion{Fe}{i} and \ion{Fe}{ii} lines:
the abundance from ionised lines is higher by about 0.1\,dex (in the case of V 0.2\,dex) then that from neutral lines.
Ni and Co (from 10 and 36 neutral lines, respectively) provide consistent values with Fe (see Fig.\,\ref{fig:alphafe}).
The \ion{Cu}{i} line at 578.2\,nm provides a negative [Cu/Fe] ratio ($\rm [Cu/Fe]=-0.37$\,dex),
but this is known to be a NLTE effect \citep[see][]{PristineXIX}.
The star is slightly enhanced in Zn ($\rm [Zn/Fe]=0.19$\,dex)
and in Zr ($\rm [Zr/Fe]=0.34$\,dex).
For Zn, according to \citet{sitnova_zn}, the NLTE correction on A(Zn) is 0.16\,dex, so taking into account NLTE corrections
on both elements, we derive $\rm [Zn/Fe]=0.24$.

We investigated two \ion{Sr}{ii} (407.7 and 421.5\,nm) lines and one \ion{Sr}{i} (460.7\,nm) to derive the Sr abundance.
The NLTE correction of the \ion{Sr}{i} is about 0.4\,dex according to \citet{bergemann12sr}.
We fitted the five \ion{Ba}{ii} lines (455.4, 493.4, 585.3, 614.1, and 649,6\,nm) available in the wavelength range 
and we derived that the star is slightly enhanced in Ba ($\rm [Ba/Fe]=0.43$) (see Fig.\,\ref{fig:alphafe}).
We verified that the partition function for La, Ce, Nd, Sm, Eu, Dy, and Er in SYNTHE was good by computing few lines with Turbospectrum and derived very consistent results.
For the heavy elements we investigated the \ion{La}{ii}, \ion{Ce}{ii}, \ion{Nd}{ii}, \ion{Sm}{ii}, \ion{Dy}{ii}, and \ion{Er}{ii} lines.
The two \ion{Eu}{ii} lines (412.9 and 664.5\,nm) were fit  and they provided [Eu/Fe]=0.25.

%%%%%%%%%%%%%%%%%%%%%%%%%%%
\subsection{Uncertainties}

The uncertainties in the stellar parameters are related to the uncertainties in the Gaia photometry and astrometry,
the way to derive the stellar parameters, and the reddening.
In Sect.\,\ref{sec:param}, different ways to derive \teff\ are discussed, each of which brings very consistent values.
Had we used the calibration by \citet{mucciarelli21} instead of the comparison to synthetic photometry and colour
as applied in Paper\,II, we would have derived a temperature 60\,K hotter.
We then assume an uncertainty of 100\,K in \teff.
The microturbulence derived from the \ion{Fe}{i} lines provides a value of 1.65\,\kms,\ while the calibration by 
\citet{mashonkina17} provides a value about 0.2\,\kms\ higher. We assigned 0.2\,\kms\ as the uncertainty in microturbulence.

For the surface gravity things are more complicated.
The uncertainty in the parallax is of 37\%, providing an uncertainty in \logg\ of 0.2\,dex.
However, deriving the \logg\ from the balance of A(Fe) from the \ion{Fe}{i} and \ion{Fe}{ii} lines, we converge to 
a \logg\ of 2.1\,dex, which is 0.3\,dex higher than the value derived from the parallax.
With \logg=1.8\,dex as derived from the Gaia\,DR3 photometry and parallax corrected by the zero point,
we obtained a perfect balance of A(Fe) from the  \ion{Fe}{i} and \ion{Fe}{ii} lines in LTE, and the value would  
still be compatible after applying the NLTE correction within the uncertainties.
With \logg=2.5\,dex we derived $\rm A(Fe)=-1.81\pm 0.11$ from the  \ion{Fe}{i} lines to which we have to add 0.08\,dex
to take into account NLTE corrections and
$\rm A(Fe)=-1.52\pm 0.12$ from the \ion{Fe}{ii} lines. The two values, $-1.73\pm 0.11$ and $-1.52\pm 0.12$\,dex, are compatible within the uncertainties. 
We adopted 0.4\,dex as the uncertainty in the surface gravity (see Fig.\,\ref{fig:loggfe}).

The uncertainties in the stellar parameters implies uncertainties in the abundances derived, and  are provided in Table\,\ref{tab:sist}.
To derive the random uncertainties we compared the results of the two UVES spectra.

%%%%%%%%%%%%%%%%%%%%%%%%%%%%%%%%%%%%%%%%%%%%%%%%%%%%%%%%%%%%%%
\section{Discussion and conclusions}

The analysis of the UVES high-resolution spectra of GHS143 has allowed us to
gain further insight into the nature of this star, and also to highlight some
difficulties in the interpretation of the data.
The first important result is that the star is a single-spectrum
spectroscopic binary (SB1), on the basis of the radial velocities measured
from the two spectra.  
As discussed in Sect.\,\ref{sec:kine}, the iron ionisation equilibrium
allows a range of possible surface gravities. The range $2.2\leq\logg \leq 2.5$
implies a distance that makes the star bound. The young apparent age and high
mass of the star remain true for all but the extreme gravity
of $\logg = 2.5$\,dex and the assumption that it is on the AGB. Even in this
extreme case, however, the age is  only
8.4\,Ga, which is younger than the bulk of the halo stars. This, coupled with the
large apogalacticon distances implied in all cases that make the star
bound, makes it more likely that the star   formed in an external galaxy and was accreted by the Milky Way.

The abundances derived for this star seem well in line with the
typical abundance patterns found in the Milky Way halo for all
elements except CNO. 
The pattern of the CNO abundances is quite exceptional.
%In Fig.\,\ref{fig:fehcno} it is compared to the unmixed stars of the sample of \citet{cayrel04}, and the 
%stars in \citet{roederer14} are surely unmixed because dwarf, turn-off, and sub-giant stars, are characterised by a $\rm [C/O]<-0.5$.
%As one can see, GHS143 is not alone, but still quite exceptional.

%%%%%%%%%%%%%%%%%%%%%%%%%%%%%%%%%%%%%%%%%%%%%%%%%%%%%%%%%%%%%%%
%\begin{figure}
%\centering
%\includegraphics[width=\hsize,clip=true]{plot_CO.pdf}
%\caption{[C/O] vs [Fe/H] for GHS143 (blue dot), 
%compared to the unmixed stars from \citet[][cyan dots]{cayrel04}
%and \citet[][red dots]{roederer14}.}
%\label{fig:fehcno}
%\end{figure}
%%%%%%%%%%%%%%%%%%%%%%%%%%%%%%%%%%%%%%%%%%%%%%%%%%%%%%%%%%%%%%%

A carbon-to-iron ratio [C/Fe] $\approx -0.4$ is compatible with what 
is observed in some ultra-faint dwarf galaxies such as Boo\,I
\citep[see e.g.][and references therein]{frebel16,norris10},
Segue\,1 \citep{norris10}, Uma\,II, and Coma Ber \citep{frebel10},
and also in the Milky Way halo \citep[see e.g.][]{barklem05}.
However, a carbon-to-oxygen ratio
[C/O]=--1.2 is far lower than what is observed in Milky Way stars \citep{akerman04,spite05}.
There are unfortunately not enough oxygen measurements
in dwarf galaxies to say much about the C/O ratio.
The ratio [(C+N+O)/Fe]=+0.65 is very high and quite exceptional.
While it may be tempting to interpret the low [C/O] ratio, coupled with the
high $\rm [N/O]=-0.03$, as the result of CNO processing, this is not possible in view
of the robustly established high $^{12}$C/$^{13}$C ratio.
The N/O ratio observed in this star is compatible
with the ratio observed in Galactic \ion{H}{II} regions
at galactocentric distances of 10--14 Kpc \citep{arellano21}, although both
the nitrogen and oxygen abundances are almost 1\,dex lower.
We propose that GHS143, whether bound or unbound, was formed
in an external galaxy. We think that the peculiar CNO abundance pattern of GHS143
is a specific signature of this galaxy, although we cannot point to
any example of a galaxy with such a chemical pattern.

%%%%%%%%%%%%%%%%%%%%%%%%%%%%%%%
%\section{Conclusions}

%%%%%%%%%%%%%%%%%%%%%%%%%%%%%%%

%%%%%%%%%%%%%%%%%%%%%%%%%%%%%%%%%%%%%%%%%%%%%%%%%%%%%%%%%%%%%%%%%%%%%%%%%%%%%%%%%%%%%%%%%%%%%%%%%%%%%%%%%%%%
\begin{acknowledgements}
The authors wish to thank the referee.
We gratefully acknowledge support from the French National Research Agency (ANR) funded projects ``Pristine'' (ANR-18-CE31-0017).
This work has made use of data from the European Space Agency (ESA) mission
{\it Gaia} (\url{https://www.cosmos.esa.int/gaia}), processed by the {\it Gaia}
Data Processing and Analysis Consortium (DPAC,
\url{https://www.cosmos.esa.int/web/gaia/dpac/consortium}). Funding for the DPAC
has been provided by national institutions, in particular the institutions
participating in the {\it Gaia} Multilateral Agreement.
This research has made use of the SIMBAD database, operated at CDS, Strasbourg, France.
\end{acknowledgements}


%-------------------------------------------------------------------



%


   \bibliographystyle{aa} % style aa.bst

   \bibliography{biblio} % your references Yourfile.bib

%


%-------------------------------------------------------------------BIBLIO-----------
%%%%%%%%%%%%%%%%%%%%%%%%%%%%%%%%%%%%%%%%%%%%%%%%%%%%%%%%%%%%%%%%%%%%%%%
\begin{appendix}

\section{Abundances}

In Table\,\ref{tab:abbo} the abundances derived from the UVES spectrum observed on November 2023 are listed.
In Table\,\ref{tab:sist} the uncertainties are reported. The second column reports the
random uncertainty related to the S/N, derived as the difference between the abundances derived from the two UVES spectra,
and the retained lines.

\begin{figure}
\centering
\includegraphics[width=\hsize,clip=true]{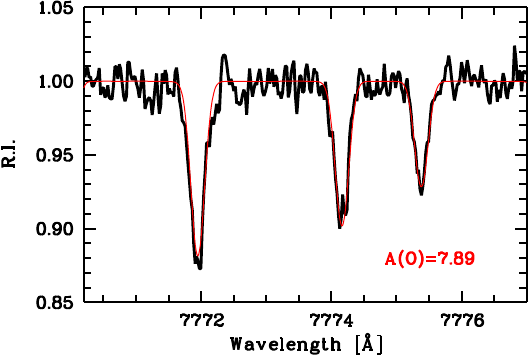}
\caption{\ion{O}{i} triplet in the observed spectrum of November 2023 (solid blck) compared to a synthesys
(solid red) with A(O) value average from the \ion{O}{i} triplet lines.}
\label{fig:o777}
\end{figure}

\begin{figure}
\centering
\includegraphics[width=\hsize,clip=true]{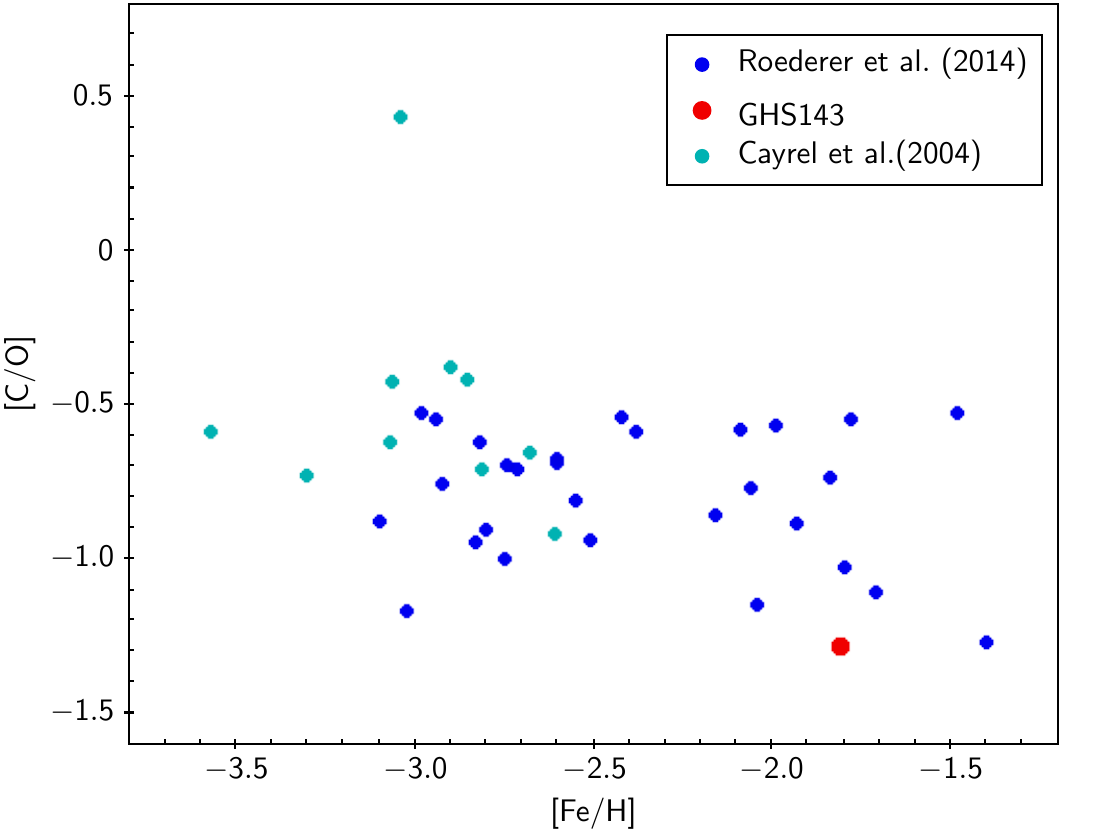}
\caption{[C/O] vs [Fe/H] for GHS143 (blue dot), 
compared to the unmixed stars from \citet[][cyan dots]{cayrel04}
and \citet[][red dots]{roederer14}.}
\label{fig:fehcno}
\end{figure}

\begin{figure}
\centering
\includegraphics[width=\hsize,clip=true]{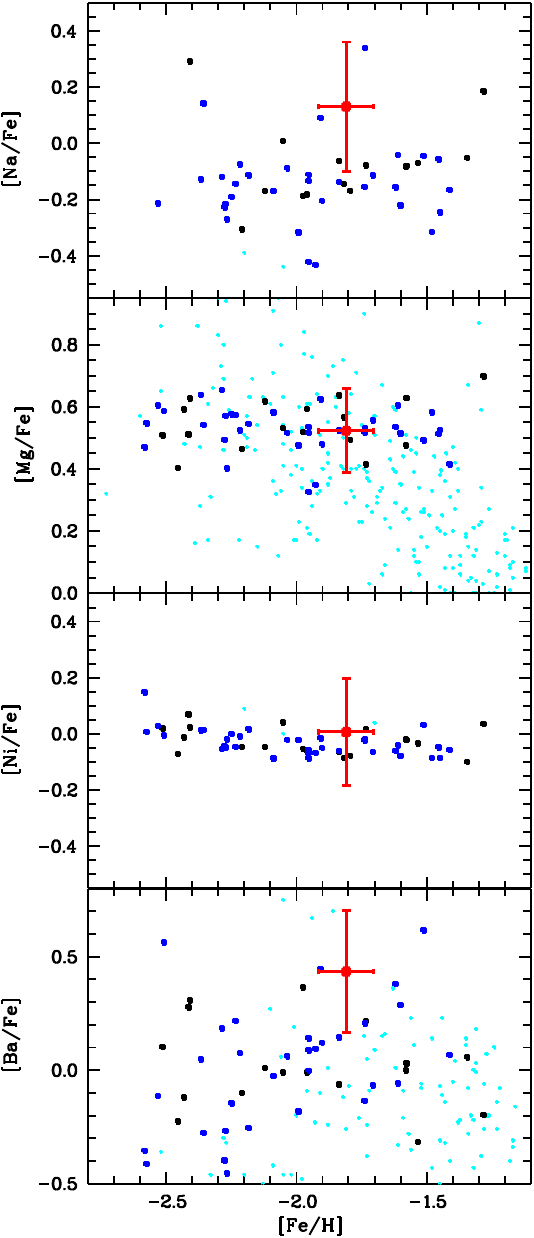}
\caption{[X/Fe] vs [Fe/H] for Na, Mg, Ni, and Ba for GHS143 (red dots), 
compared to the sample from \citet[][blue dots]{PristineXIX}
and \citet[][black dots]{mince1}, and from stars from the  Sculptor dwarf galaxy taken from the SAGA database
\citep[][light blue dots]{SAGA1,SAGA2}.}
\label{fig:alphafe}
\end{figure}

In Fig.\,\ref{fig:o777} the oxygen triplet is shown.
In Fig.\,\ref{fig:fehcno} [C/O] vs [Fe/H] of GHS143 is compared to the values of the unmixed stars of the sample of \citet{cayrel04}, and the 
stars in \citet{roederer14} characterised by a $\rm [C/O]<-0.5$ that are surely unmixed because dwarf, turn-off, and sub-giant stars.
As one can see, GHS143 is not alone, but still quite exceptional.
In Fig.\,\ref{fig:alphafe} the [X/Fe] vs [Fe/H] plots for Na, Mg, Ni and Ba are shown.

%%%%%%%%%%%%%%%%%%%%%%%%%%%%%%%%%%%%%%%%%%%%%%%%%%%%%%%%%%%%%%
%%%%%%%%%%%%%%%%%%%%%%%%%%%%%%%%%%%%%%%%%%%%%%%%%%%%%%%%%%%%%%

%%%%%%%%%%%%%%%%%%%%%%%%%%%%%%%%%%%%%%%%%%%%%%%%%%%%%%%%%%%%%%
%%%%%%%%%%%%%%%%%%%%%%%%%%%%%%%%%%%%%%%%%%%%%%%%%%%%%%%%%%%%%%

%%%%%%%%%%%%%%%%%%%%%%%%%%%%%%%%%%%%%%%%%%%%%%%%%%%%%%%%%%%%%%
%%%%%%%%%%%%%%%%%%%%%%%%%%%%%%%%%%%%%%%%%%%%%%%%%%%%%%%%%%%%%%

%%%%%%%%%%%%%%%%%%%%%%%%%%%%%
\begin{table*}
\caption{Abundances.}
\label{tab:abbo}
\begin{tabular}{lrrrrrrrr}
\hline
\smallskip
Element & Nlin & $\rm A(X)\odot$ & A(X) & [X/H] & $\sigma$ & [X/Fe] & $\sigma$ & Cor(NLTE) \\
\hline
\ion{C}{i}   &   1 &  8.50 & $ 6.300$ & $-2.200$ & 0.100 & $-0.390$ & 0.145 &  \\
\ion{N}{i}   &   1 &  7.86 & $ 6.831$ & $-1.029$ & 0.100 & $ 0.779$ & 0.145 &  \\
\ion{O}{i}   &   4 &  8.76 & $ 7.849$ & $-0.911$ & 0.085 & $ 0.897$ & 0.135 & $-0.11$ \\
\ion{Na}{i}  &   4 &  6.30 & $ 4.623$ & $-1.677$ & 0.204 & $ 0.131$ & 0.230 & $-0.20$ \\
\ion{Mg}{i}  &  10 &  7.54 & $ 6.255$ & $-1.285$ & 0.084 & $ 0.523$ & 0.135 & $0.08$ \\
\ion{Al}{i}  &   1 &  6.47 & $ 4.311$ & $-2.159$ &       & $-0.351$ &       & $0.65$ \\
\ion{Si}{i}  &  15 &  7.52 & $ 6.170$ & $-1.350$ & 0.089 & $ 0.458$ & 0.138 & $-0.03$ \\
\ion{Si}{ii} &   1 &  7.52 & $ 6.171$ & $-1.349$ &       & $ 0.354$ &       &  \\
\ion{K}{i}   &   1 &  5.11 & $ 4.236$ & $-0.874$ &       & $ 0.934$ &       & $0.37$ \\
\ion{Ca}{i}  &  23 &  6.33 & $ 4.994$ & $-1.336$ & 0.112 & $ 0.473$ & 0.154 & $0.09$ \\
\ion{Sc}{i}  &   1 &  3.10 & $ 1.261$ & $-1.839$ &       & $-0.030$ &       &  \\
\ion{Sc}{ii} &   4 &  3.10 & $ 1.517$ & $-1.583$ & 0.061 & $ 0.226$ & 0.122 &  \\
\ion{Ti}{i}  &  23 &  4.90 & $ 3.490$ & $-1.410$ & 0.135 & $ 0.398$ & 0.171 &  \\
\ion{Ti}{ii} &  32 &  4.90 & $ 3.604$ & $-1.296$ & 0.080 & $ 0.407$ & 0.135 &  \\
\ion{V}{i}   &   9 &  4.00 & $ 2.221$ & $-1.779$ & 0.137 & $ 0.029$ & 0.173 &  \\
\ion{V}{ii}  &   5 &  4.00 & $ 2.454$ & $-1.546$ & 0.084 & $ 0.157$ & 0.137 &  \\
\ion{Cr}{i}  &  16 &  5.64 & $ 3.810$ & $-1.830$ & 0.106 & $-0.022$ & 0.149 &  \\
\ion{Cr}{ii} &   4 &  5.64 & $ 3.898$ & $-1.742$ & 0.120 & $-0.040$ & 0.161 &  \\
\ion{Mn}{i}  &  13 &  5.37 & $ 3.214$ & $-2.156$ & 0.047 & $-0.347$ & 0.116 &  \\
\ion{Mn}{ii} &   1 &  5.37 & $ 3.352$ & $-2.018$ &       & $-0.315$ &       &  \\
\ion{Fe}{i}  & 244 &  7.52 & $ 5.712$ & $-1.808$ & 0.106 & $ 0.000$ &       & $0.11$ \\
\ion{Fe}{ii} &  23 &  7.52 & $ 5.817$ & $-1.703$ & 0.108 & $ 0.000$ &       &  \\
\ion{Co}{i}  &  10 &  4.92 & $ 3.150$ & $-1.770$ & 0.145 & $ 0.039$ & 0.180 &  \\
\ion{Ni}{i}  &  36 &  6.23 & $ 4.429$ & $-1.801$ & 0.159 & $ 0.007$ & 0.191 &  \\
\ion{Cu}{i}  &   1 &  4.21 & $ 2.030$ & $-2.180$ &       & $-0.371$ &       &  \\
\ion{Zn}{i}  &   2 &  4.62 & $ 3.004$ & $-1.616$ & 0.005 & $ 0.192$ & 0.106 &  \\
\ion{Sr}{i}  &   1 &  2.92 & $ 1.043$ & $-1.877$ &       & $-0.069$ &       & 0.4  \\
\ion{Sr}{ii} &   2 &  2.92 & $ 1.476$ & $-1.444$ & 0.078 & $ 0.259$ & 0.134 &  \\
\ion{Y}{ii}  &   9 &  2.21 & $ 0.475$ & $-1.735$ & 0.145 & $-0.032$ & 0.181 &  \\
\ion{Zr}{ii} &   9 &  2.62 & $ 1.237$ & $-1.383$ & 0.136 & $ 0.320$ & 0.174 &  \\
\ion{Ba}{ii} &   5 &  2.17 & $ 0.893$ & $-1.277$ & 0.248 & $ 0.435$ & 0.270 &  \\
\ion{La}{ii} &  10 &  1.14 & $-0.524$ & $-1.664$ & 0.076 & $ 0.039$ & 0.132 &  \\
\ion{Ce}{ii} &  10 &  1.61 & $-0.145$ & $-1.755$ & 0.079 & $-0.052$ & 0.134 &  \\
\ion{Nd}{ii} &  21 &  1.45 & $-0.261$ & $-1.711$ & 0.134 & $-0.008$ & 0.172 &  \\
\ion{Sm}{ii} &   6 &  1.00 & $-0.461$ & $-1.461$ & 0.099 & $ 0.242$ & 0.146 &  \\
\ion{Eu}{ii} &   2 &  0.52 & $-0.935$ & $-1.455$ & 0.005 & $ 0.248$ & 0.109 &  \\
\ion{Dy}{ii} &   1 &  1.13 & $-0.405$ & $-1.535$ &       & $ 0.168$ &       &  \\
\ion{Er}{ii} &   1 &  0.96 & $-0.847$ & $-1.807$ &       & $-0.104$ &       &  \\
\hline
\end{tabular}
\end{table*}
%%%%%%%%%%%%%%%%%%%%%%%%%%%%%

%%%%%%%%%%%%%%%%%%%%%%%%%%%%%
\begin{table*}
\caption{Uncertanties on [X/Fe].}
\label{tab:sist}
\begin{tabular}{lrrrrr}
\hline
\smallskip
Ellement  & [X/H]  & Nlines &  \teff\      & \logg\         & $\xi$           \\
          & random &        & $\pm 100$\,K & $\pm 0.4$\,dex & $\pm 0.2$\,\kms \\
\hline
\ion{Fe}{i}  & 0.00 & 244/254 & $0.10$       & $0.01$ & $0.03$         \\
\ion{Fe}{ii} & 0.00 &  23/22  & $0.01$       & $0.18$ & $0.04$         \\
\ion{C}{i}   & 0.01 &  G-band & $0.08$       & $0.11$ & $0.00$         \\
\ion{O}{i}   & 0.03 &   4/4   & $0.17$       & $0.18$ & $0.02$         \\
\ion{Na}{i}  & 0.03 &   4/4   & $0.04$       & $0.00$ & $0.01$         \\
\ion{Mg}{i}  & 0.05 &   8/10  & $0.04$       & $0.05$ & $0.00$         \\
\ion{Al}{i}  & 0.02 &   1/1   & $0.10$       & $0.15$ & $0.00$         \\
\ion{Si}{i}  & 0.01 &  15/12  & $0.06$       & $0.02$ & $0.03$         \\
\ion{Si}{ii} & 0.02 &   1/1   & $0.01$       & $0.09$ & $0.02$         \\
\ion{K}{i}   & 0.05 &   1/1   & $0.00$       & $0.08$ & $0.04$         \\
\ion{Ca}{i}  & 0.01 &  26/23  & $0.02$       & $0.00$ & $0.01$         \\
\ion{Sc}{i}  & 0.11 &   1/1   & $0.04$       & $0.01$ & $0.02$         \\
\ion{Sc}{ii} & 0.02 &   4/3   & $0.06$       & $0.10$ & $0.00$         \\
\ion{Ti}{i}  & 0.02 &  23/28  & $0.00$       & $0.01$ & $0.01$         \\
\ion{Ti}{ii} & 0.00 &  32/35  & $0.04$       & $0.01$ & $0.01$         \\
\ion{V}{i}   & 0.01 &   9/8   & $0.05$       & $0.01$ & $0.02$         \\
\ion{V}{ii}  & 0.04 &   5/4   & $0.04$       & $0.06$ & $0.00$         \\
\ion{Cr}{i}  & 0.01 &  16/14  & $0.00$       & $0.00$ & $0.02$         \\
\ion{Cr}{ii} & 0.02 &   4/4   & $0.01$       & $0.02$ & $0.02$         \\
\ion{Mn}{i}  & 0.02 &  13/15  & $0.01$       & $0.01$ & $0.01$         \\
\ion{Mn}{ii} & 0.04 &   1/1   & $0.08$       & $0.10$ & $0.03$         \\
\ion{Co}{i}  & 0.04 &  10/9   & $0.03$       & $0.00$ & $0.01$         \\
\ion{Ni}{i}  & 0.02 &  36/32  & $0.01$       & $0.00$ & $0.02$         \\
\ion{Cu}{i}  &      &   1/0   & $0.06$       & $0.00$ & $0.03$         \\
\ion{Zn}{i}  & 0.04 &   2/2   & $0.06$       & $0.08$ & $0.01$         \\
\ion{Sr}{ii} & 0.04 &   2/2   & $0.10$       & $0.15$ & $0.00$         \\
\ion{Y}{ii}  & 0.03 &  9/11   & $0.06$       & $0.01$ & $0.02$         \\
\ion{Zr}{ii} & 0.01 &   9/9   & $0.07$       & $0.06$ & $0.00$         \\
\ion{Ba}{ii} & 0.04 &  5/3    & $0.06$       & $0.02$ & $0.11$         \\
\ion{La}{ii} & 0.04 & 10/10   & $0.07$       & $0.02$ & $0.03$         \\
\ion{Ce}{ii} & 0.02 & 11/15   & $0.07$       & $0.05$ & $0.04$         \\
\ion{Nd}{ii} & 0.02 & 21/21   & $0.05$       & $0.06$ & $0.03$         \\
\ion{Sm}{ii} & 0.10 &  7/6    & $0.07$       & $0.01$ & $0.04$         \\
\ion{Eu}{ii} & 0.08 &  2/2    & $0.04$       & $0.04$ & $0.04$         \\
\ion{Dy}{ii} & 0.03 &  1/1    & $0.09$       & $0.07$ & $0.06$         \\
\ion{Er}{ii} & 0.07 &  1/1    & $0.06$       & $0.01$ & $0.03$         \\
\hline
\end{tabular}
\end{table*}
%%%%%%%%%%%%%%%%%%%%%%%%%%%%%

%%%%%%%%%%%%%%%%%%%%%%%%

%%%%%%%%%%%%%%%%%%%%%%%%%%%%%%%%%%%%%%%%%%%%%%%%%%%%%
\end{appendix}

\end{document}